\documentclass[twocolumn,twoside]{IEEEtran}
\usepackage{times}
\usepackage{amsmath}
\usepackage{amsbsy}
\usepackage{latexsym}
\usepackage[ansinew]{inputenc}
\usepackage[dvips ]{graphicx}
\input epsf\title{{Adaptive Space-Time Decision Feedback Neural
Detectors with Data Selection for High-Data Rate Users in DS-CDMA
Systems}}
\author{Rodrigo C. de Lamare and Raimundo Sampaio-Neto  \\
\thanks{\footnotesize Dr. R. C. de Lamare is a Lecturer with the Communications Research Group,
Department of Electronics, University of York, York Y010 5DD,
United Kingdom and Prof. R. Sampaio-Neto is with CETUC/PUC-RIO,
22453-900, Rio de Janeiro, Brazil. E-mails:
rcdl500@ohm.york.ac.uk, raimundo@cetuc.puc-rio.br} }

\date{  }
\begin{document}
\maketitle
\begin{abstract}
A space-time adaptive decision feedback (DF) receiver using
recurrent neural networks (RNN) is proposed for joint equalization
and interference suppression in direct-sequence
code-division-multiple-access (DS-CDMA) systems equipped with
antenna arrays. The proposed receiver structure employs
dynamically driven RNNs in the feedforward section for
equalization and multi-access interference suppression and a
finite impulse response (FIR) linear filter in the feedback
section for performing interference cancellation. A data selective
gradient algorithm, based upon the set-membership design
framework, is proposed for the estimation of the coefficients of
RNN structures and is applied to the estimation of the parameters
of the proposed neural receiver structure. Simulation results show
that the proposed techniques achieve significant performance gains
over
existing schemes.\\


\end{abstract}

\section{Introduction}

\PARstart{C}{ode} division multiple access (CDMA) implemented with
direct sequence (DS) spread-spectrum signalling is currently
deployed in many wireless communication systems. Such services
include third-generation (3G) cellular telephony, indoor wireless
networks, terrestrial and satellite communication systems. The
advantages of CDMA include good performance in multi-path
channels, flexibility in the allocation of channels, increased
capacity in bursty and fading environments and the ability to
share bandwidth with narrowband communication systems without
deterioration of either's systems performance \cite{honig,verdu}.

The capacity of a DS-CDMA network is limited by different types of
interference, namely, narrowband interference (NBI), multi-access
interference (MAI), inter-symbol interference (ISI) and the noise
at the receiver. The major source of interference in most CDMA
systems is MAI, which is highly structured and arises due to the
fact that users communicate through the same physical channel with
non-orthogonal signals. The multiple propagation paths of the
channel can destroy the orthogonality of the signals at the
receiver, giving rise to MAI even when the users employ orthogonal
signatures. For non-orthogonal signatures the channel and the lack
of synchronism between users can further increase the level of
MAI. The conventional (single-user) receiver that employs a filter
matched to the signature sequence does not suppress MAI and is
very sensitive to differences in power between the received
signals (near-far problem). In order to mitigate the effect of
interference in CDMA systems, a designer can resort to multiuser
detection techniques for MAI suppression, and antenna arrays,
which perform spatial filtering and can reduce the impact of
interference. The optimal multiuser detector of Verdu
\cite{verdu86} suffers from exponential complexity and requires
the knowledge of timing, amplitude and signature sequences. This
has motivated the development of various sub-optimal strategies:
the linear \cite{lupas} and decision feedback \cite{falconer}
receivers, the successive interference canceller \cite{patel} and
the multistage detector \cite{varanasi}. The combination of
multiuser detection and beamforming can provide an enhanced
performance for MAI and ISI suppression \cite{honig,verdu}. This
requires the joint processing of the data received at an antenna
array with elements closely spaced and the estimation of the
direction of arrival (DoA) and the channel information.

In current third-generation (3G) wireless cellular communications,
there are some specific features in the operation of the systems
such as the use of antenna arrays by the base stations for
increasing the capacity in the uplink (mobile terminals to base
station) and different data rates and quality of service
requirements \cite{3G}. In particular, third-generation wideband
DS-CDMA systems have multi-code capabilities, in which the user
signals with high data rate (HD) can be accommodated by reducing
the processing gain $N_h$ and using a low spreading factor and
higher power levels \cite{3G}. In these scenarios, users are
equipped with multiple processing gains, operate at different data
rates, and HD users operate at higher signal-to-noise ratios
(SNRs). The SNR is proportional to their data rates, which leads
to a situation where they experience a relatively low level of MAI
but more significant level of ISI. In contrast, low data rate (LD)
users, that work with larger spreading factors and lower SNRs,
encounter higher levels of MAI and near-far effects.

In order to design an effective space-time receiver for
interference mitigation in DS-CDMA systems employing multi-code
capabilities and antenna arrays, we investigate in this work the
use of receiver structures based on recurrent neural networks and
decision feedback schemes. One motivation for this work is that
neural networks provide good nonlinear mapping of the inverse
model of the channel and can deal with the uncertainty and
interference present in the received data. Therefore, receivers
based on neural networks outperform those based on FIR linear
filters. Despite the increased complexity over receivers with FIR
linear filters, the deployment of neural structures is feasible in
situations where the number of processing elements is not very
large, i.e., the spreading factor of CDMA systems is low and the
number of HD users is small. Such situations appear in the context
of 3G wireless communications \cite{3G}. Recent work on
equalization and modelling of nonlinear systems has shown that
recurrent neural networks (RNNs) are superior in bit error rate
(BER) performance to multilayer perceptrons (MLPs), radial basis
functions (RBFs) \cite{haykin-neural} and FIR linear filters.
Prior work on equalization structures also demonstrated that
decision feedback (DF) neural equalization structures are better
than neural schemes without DF and DF and linear equalizers with
linear FIR filters \cite{ong}. In the context of CDMA systems,
existing works on neural and related structures are restricted to
MLPs \cite{aazhang}, RBFs \cite{mitra}, support vector machines
(SVMs) \cite{hanzo} and to static channels, and little attention
has been given to RNNs and DF structures.  This is especially
relevant for the uplink of wireless communications systems, where
the base station can afford an antenna array and more complex
algorithms as compared to the mobile terminals. In this case, the
trade-off between computational complexity and superior
performance is quite attractive.

In order to estimate the parameters of the receiver, a designer
must resort to a training algorithm. Amongst the available
training methods for neural detectors, one can employ gradient
based approaches such as the back propagation algorithm for MLPs
\cite{haykin-neural} and the real time recurrent learning (RTRL)
or fast least-squares (LS) and Kalman-based techniques \cite{choi}
for RNNs. On one hand, gradient-type approaches for neural schemes
are simpler to implement, exhibit low complexity but possess slow
convergence rates and have to deal with the problem of vanishing
gradients \cite{haykin-neural} for RNNs. On the other hand, LS and
Kalman algorithms have fast convergence but exhibit very high
computational complexity and numerical problems for
implementation. This motivates the search for novel approaches to
training RNNs that possess fast convergence, low complexity and
flexibility. The algorithm can be also applied to problems such as
estimation and tracking of non-linear and dynamic processes and
modelling of time-varying non-linear systems \cite{haykin-neural}.
In this context, set-membership (SM) filtering
\cite{huang,gollamudi} represents a class of recursive estimation
algorithms that, on the basis of a pre-determined error bound,
seeks a set of parameters that yield bounded output errors. In
particular, the SM approach leads to low-complexity gradient
algorithms with excellent convergence and tracking performance due
to the use of an adaptive step size and data selective updates.
The appealing feature of SM algorithms is that they are able to
significantly improve the convergence and tracking of estimators
and possess a complexity lower than conventional estimation
algorithms due to their sparse updates. The existing works on SM
algorithms deal with the adaptation of linear FIR filters and
there is no SM algorithm for training important neural structures
such as RNNs available.

In this work, our first contribution is to present an adaptive
space-time adaptive DF multiuser receiver structure for HD users
with antenna arrays, using dynamically driven RNNs
\cite{haykin-neural} in the feedforward section and FIR linear
filters in the feedback section for cancelling the associated HD
and LD users. These RNNs have not been considered for the problem
of space-time multiuser detection so far. The advantage of antenna
arrays lies in the spatial separation of the interfering users,
increasing the capacity of the system. The proposed structure
employs antenna arrays combined with a neural RNN processor and a
DF structure, which are shown to provide considerable capacity
gains over schemes with FIR linear filters. The second
contribution of this work is the development of a novel
data-selective algorithm for parameter estimation in RNNs using
the set-membership estimation framework \cite{huang,gollamudi}.
The proposed algorithm is applied here to the problem of
space-time neural receivers for DS-CDMA systems, even though, the
algorithm is general and can be used for other applications of
neural networks. It should also be remarked that some very recent
contributions \cite{xia}-\cite{delgado}, which constitute the
state-of-the-art in the area of RNN, can be considered for further
investigation in the context of interference mitigation in
communication systems such as CDMA, OFDM and MIMO \cite{paulraj}.
However, we have opted to focus here on a RNN scheme combined with
FIR filters in a decision feedback architecture, its application
to space-time CDMA systems and the development of a novel
estimation technique.

This paper is structured as follows. Section II briefly describes
the space-time DS-CDMA communication system model and formulates
the problem. The proposed space-time DF neural receivers for joint
equalization and multiuser detection are presented in Section III.
The novel set-membership adaptive algorithms for the neural
structures are presented in Section IV. Section V is dedicated to
the presentation and discussion of the simulation results and
Section VI gives the concluding remarks of this work and comments
on possible extensions and future work.

\section{Space-time DS-CDMA system model and problem statement}

Let us consider the uplink of an asynchronous multi-code QPSK
DS-CDMA system with $L$ paths and $2$ classes of $K=K_h + K_l$
users, where $K_h$ is the number of HD users with $N_h$ chips per
symbol and $K_l$ the number of LD users with $N_l$ chips per
symbol, as depicted in Fig. 1. These QPSK symbols are spread with
unique sequences for each user, modulated and transmitted over a
communication channel. At the receiver equipped with antenna
arrays, the composite signal is demodulated, sampled at chip rate
and applied to a bank of filters matched to the users´ effective
spatial spreading sequences. In what follows, we mathematically
detail these operations and describe the signal processing.

The transmitted signal for the $k$th user amongst the HD users and
the $q$th user amongst the LD users are given by
\begin{equation}
\begin{split}
x_{k}^h(t)=A_{k}^h\sum_{i=-\infty}^{\infty}b_{k}^h(i)s_{k}^h(t-iT_h)\\
x_{q}^l(t)=A_{q}^l\sum_{i=-\infty}^{\infty}b_{k}^l(i)s_{q}^l(t-iT_l)
\end{split}
\end{equation}
where the equiprobable symbols $b_{k}^h(i) ~\textrm{and}~ b_q^l(i)
\in \{\pm1, \pm j\}/ \sqrt{2}$ with $j^2=-1$ denote the $i$-th
symbol for users $k$ and $q$, the real valued spreading waveforms
and the amplitude associated with HD user $k$ and LD user $q$ are
$s_{k}^h(t)$, $s_q^l(t)$, $A_k^h$ and $A_{q}^l$, respectively.

The spreading waveforms for the HD and LD users are expressed by
$s_{k}^h(t) = \sum_{i=1}^{N_h} a_{k}^h(i) \phi(t-iT_{c})$ and
$s_{k}^l(t) = \sum_{i=1}^{N_l} a_{k}^l(i) \phi(t-iT_{c})$,
respectively, where $a_{k}^h(i)\in \{\pm1/\sqrt{N_h} \}$,
$a_{k}^l(i)\in \{\pm1/\sqrt{N_l} \}$ , $\phi(t)$ is the chip
waveform, $T_{c}$ is the chip duration and $N_h=T_h/T_{c}$ and
$N_l=T_l/T_{c}$ are the processing gains of the HD and LD users,
respectively. In practice, HD users can use $N_h = \beta N_l$,
being $\beta = 2, 4, 8, 16, 32$, the smallest $N_l =8$ and the
largest $N_h=256$ \cite{umts}. Assuming that the receiver with
linear antenna arrays is synchronized with the main path and
identical fading is experienced by all antenna elements for each
path of each user signal, the coherently demodulated composite
received signal at the $p$th antenna element is
\begin{equation}
\begin{split}
r_p(t) & = \sum_{k=1}^{K_h}\sum_{m=0}^{L-1} h_{k,m}^h(t)e^{-j
\Theta_{k,m}^h} x_{k}^h(t-\tau_{k,m}^h-d_k^h) \\ & \quad +
\sum_{q=1}^{K_l}\sum_{m=0}^{L-1} h_{q,m}^l(t)e^{-j \Theta_{q,m}^l}
x_{q}^l(t-\tau_{q,m}^l-d_q^l) +n(t)
\end{split}
\end{equation}
where $\Theta_{k,m}^h=2 \pi (p-1)(d/\lambda) \sin (\phi_{k,m}^h)$
and $\Theta_{q,m}^l=2 \pi (p-1)(d/\lambda) \sin (\phi_{q,m}^l)$
are the delay shifts of the $m$th path of the $k$th HD user and
the $q$th LD user, respectively. The quantities $\phi_{k,m}^h$ and
$\phi_{q,m}^l$ are the directions of arrival (DoA) of the signals
of the HD user $k$ and LD user $q$ and their $m$th paths,
$d=\lambda/2$ is the spacing between sensors and $\lambda$ is the
carrier wavelength. The channel coefficients associated with the
$m$-th path and the $k$-th HD user are $h_{k,m}^h(t)$, $d_k^h
\in[0,~N_h)$ is the delay of the $k$th user taken from a discrete
uniform random variable between $0$ and $N_h$ and $\tau_{k,m}^h$
is the delay of the $m$th path of the $k$th user, whereas the
channel coefficients associated with the $m$-th path and the
$q$-th LD user are $h_{q,m}^l(t)$, $d_q^l \in[0,~N_l)$ is the
delay of the $q$th user taken from a discrete uniform random
variable between and $N_l$ $0$ and $\tau_{q,m}^l$ is the delay of
the $m$th path of the $q$th LD user. The coefficients of the
channels for both LD and HD users are modelled as a
tapped-delay-line and their coefficients can be computed according
to an available standard such as the UMTS model \cite{umts}.

We assume that the delays are multiples of the chip periods, the
channel is constant during each symbol interval (the coherence
time of the channel does not exceed that of the symbol - this is
typical in 3G systems \cite{3G}), the spreading codes are repeated
from symbol to symbol, and the receiver with a $J$-element linear
antenna array is perfectly synchronized with the main path. The
complex envelope of the received waveforms after filtering by a
chip-pulse matched filter and sampled at chip rate yields the
discrete-time samples for the user of interest at the $p$th
antenna element
\begin{equation}
\begin{split}
r_{p} (n) & = \sum_{k=1}^{K_h} \sum_{m=0}^{L-1} h_{k,m}^h e^{-j
\Theta_{k,m}^h} x_{k,m}^h (n T_c - \tau_{k,m}^h -d_k^h) \\ & \quad
+ \sum_{q=1}^{K_l} \sum_{m=0}^{L-1} h_{q,m}^l e^{-j
\Theta_{q,m}^l} x_{q,m}^l (n T_c - \tau_{q,m}^l -d_q^l) +  n_p (n)
\end{split}
\end{equation}
By collecting these samples from each antenna element and
organizing them into a $JM\times 1$ observation vector
corresponding to the $i$th signalling interval, we obtain
\begin{equation}
\begin{split}
{\bf r}(i) & =  \sum_{k=1}^{K_h} {\bf x}_k (i)  + \sum_{q=1}^{K_l}
{\bf z}_q(i) + \boldsymbol{\eta}(i)   + {\bf n}(i)
\end{split}
\end{equation}
where $M=N_h+L-1$, the $k$th HD user signal is given by ${\bf x}_k
(i)$, the $q$th asynchronous signal of LD users that impinges on
the antenna array is represented by ${\bf z}_q(i)$. The $(JM
\times 1)$ vector ${\bf z}_q(i)$ is given by ${\bf z}_q(i)= [
z_{q,1}(i) \ldots z_{q,JM}(i) ] ^T$, where $z_{q,n}(i) =
\sum_{m=0}^{L-1} h_{q,m}^l e^{-j \Theta_{q,m}^l} x_{q,m}^l(nT_c -
\tau_{q,m}^l - d_q^l)$.

The complex Gaussian noise vector is ${\bf n}(i) = [n_{1}(i)
~\ldots~n_{JM}(i)]^{T}$ with zero mean and $E[{\bf n}(i){\bf
n}^{H}(i)] = \sigma^{2}{\bf I}$, $(\cdot)^{T}$ and $(\cdot)^{H}$
denote transpose and Hermitian transpose, respectively, the
operator $E[\cdot]$ stands for expected value, and
$\boldsymbol{\eta}(i)$ is the intersymbol interference (ISI)
vector. Indeed, the $JM \times 1$ vector of ISI
$\boldsymbol{\eta}(i)$ is constructed with linear combinations of
the channel coefficients, the symbols from previous, current and
succeeding symbols and chips of the spreading codes of the HD
users, similarly to ${\bf z}_q(i)$. In this model, the ISI span
and contribution ${\boldsymbol \eta}[i]$ are functions of the
processing gain $N_h$ of HD users and the number of paths $L$. If
$1< L \leq N_h$ then $3$ symbols would interfere in total, the
current one, the previous and the successive symbols. In the case
of $N_h < L \leq 2N_h$ then $5$ symbols would interfere, the
current one, the $2$ previous and the $2$ successive ones. In most
practical CDMA systems, we have that $1< L \leq N_h$ and then only
$3$ symbols are usually affected. The reader is referred to
\cite{honig} for further details related to the ISI and models.
The UMTS channel models \cite{umts}, which reveal that the channel
usually affects at most $3$ symbols (it typically spans a few
chips) provides further details about the typical channels.  In
the case of HD users it is substantially more significant due to
the ratio $L/N_h$ which is greater than $L/N_l$.

The main problem is how to detect the HD users with the minimum
probability of error possible and at reasonably complexity. The
minimum probability of error detector was proposed by Verdu
\cite{verdu,verdu86} and corresponds to a combinatorial problem
with exponential complexity with the number of users. In what
follows, we seek the design of a high-performance detector with
affordable complexity by proposing a decision feedback structure
with recurrent neural networks.

\section{Proposed Space-Time Decision Feedback Neural Receiver}

The space-time decision feedback receiver structure, depicted in
Fig. 2, applies a bank of ST-RAKE detectors \cite{paulraj} to the
observation vector ${\bf r}(i)$, followed by a neural MUD that
employs dynamically driven recurrent neural networks (RNNs) in the
feedforward section for suppressing MAI and ISI and an FIR linear
filter in the feedback section for cancelling the associated users
in the system. The RNNs \cite{haykin-neural} used in the multiuser
detector (MUD) are small structures, with feedback connections and
where each artificial neuron is connected to the others, capable
of achieving superior performance to MLPs and RBFs
\cite{haykin-neural}. In this context, the advantage of the neural
MUD over linear ones is the use of non-linear mappings to create
decision boundaries for the detection of transmitted symbols.

We consider a one shot approach (detection of the symbols
corresponding to one time instant) where the $K_h\times 1$ input
vector ${\bf u}(i)=[u_{1}(i)~\ldots~u_{K_h}(i)]^{T}$ to the MUD is
given by
\begin{equation}
{\bf u}(i)={\bf S}^H{\bf r}(i),
\end{equation}
where ${\bf S} = [\tilde{\bf s}_{1}^{a}~\ldots~\tilde{\bf
s}_{K_h}^{a}]$ and $\tilde{\bf s}_{k}^{a}$ is the user $k$
effective spatial signature of the $k$th HD user that assumes the
knowledge (or estimates) of the DOAs and channels or their
estimates. The spatial signature $\tilde{\bf s}_{k}^{a}$ results
from the convolution of the original signature sequence ${\bf
s}_k^h$ with the channel and the time shifts resulting from the
antenna array, and can be represented as
\begin{equation}
\tilde{\bf s}_{k}^{a} = [ \tilde{\bf s}_{k,1}^T ~ \ldots ~
\tilde{\bf s}_{k,p}^T ~\ldots ~ \tilde{\bf s}_{k,J}^T ]^T,
\end{equation}
where $\tilde{\bf s}_{k,p} = [ \tilde{s}_{k,p,1} \ldots ~
\tilde{s}_{k,p,M}]^T$ is the $M \times 1$ signature at the $p$th
antenna of the receiver.

Let us now describe in detail the schematic of the proposed
space-time decision feedback neural receiver depicted in Fig. 2.
Specifically, the proposed receiver design is equivalent to
processing the output ${\bf u}(i)$ of the bank of ST-RAKE
detectors with the RNN in the feedforward part of the receiver and
the decision feedback section. The $K_h \times 1$ data vector
${\bf u}(i)$ is stacked with the $Q \times 1$ vector of states of
the RNN
\begin{equation}
{\bf x}_k(i-1) = \varphi({\bf W}_{k}^{H}(i)\boldsymbol{
\xi}_{k}(i)),
\end{equation}
where the matrix ${\bf W}_{k}(i)=[{\bf w}_{k,1}(i) ~\ldots~{\bf
w}_{k,j}(i)~\ldots~{\bf w}_{k,Q}(i)]$ with dimensions $(Q+K_h)
\times Q$ and whose $Q$ columns ${\bf w}_{k,j}(i)$, with $j=1,~
2,~\ldots, Q$ have dimensions $(Q+K_h)\times 1$ contain the
coefficients of the RNN receiver for user $k$ and $\varphi(.)$ is
the activation function of the RNN and which is used to form the
$(Q+K_h) \times 1$ vector
\begin{equation}
\boldsymbol{\xi}_k(i)= [{\bf x}_k^T(i-1) ~{\bf u}^T(i)]^T,
\end{equation}
which is processed by the neural part of the receiver. The
activation function is chosen as $\varphi(\cdot)={\rm tanh}
(\cdot)$ in order to limit the amplitude range of the output
signals of the network and to avoid numerical instability in the
recurrent structure \cite{haykin}. In particular, the ${\rm tanh}
(\cdot)$ fits well in communications problems as it provides
naturally a soft estimate of symbols. In the case of DS-CDMA
systems, it provides a soft estimate which is fed back to the
neural structure and helps it to inversely map the symbol estimate
appropriately in the presence of multiple-access interference and
noise. The RNN in the feedforward section then processes
$\boldsymbol{\xi}_k(i)$ to yield the initial decisions as given by
\begin{equation}
\begin{split} \hat{b}_k(i) & = v \Big( {\bf Dx}_{k}(i)
\Big)  = {\rm sgn} \Big( \Re \big[ {\bf Dx}_{k}(i) \big] \Big) +
\j {\rm sgn} \Big( \Im \big[ {\bf Dx}_{k}(i) \big] \Big)
\\ & = {\rm sgn} \Big( \Re \big[ {\bf D} \varphi({\bf
W}_{k}^{H}(i)\boldsymbol{ \xi}_{k}(i)) \big] \Big) + \j {\rm sgn}
\Big( \Im \big[ {\bf D} \varphi({\bf W}_{k}^{H}(i)\boldsymbol{
\xi}_{k}(i)) \big] \Big) ,
\end{split}
\end{equation}
where $v(\cdot)$ denotes a slicing function that contains the
operators $\Re(.)$ and $\Im(.)$ which are used to select the real
and imaginary parts, respectively, and then quantizes the output
of each operator in order to detect the QPSK symbols. The matrix
${\bf D}=\big[1~0~...~0\big]$ is the { $1\times Q$} matrix that
defines the number of outputs of the network and  and ${\rm
sgn}(.)$ is the signum function. Since we are only interested in
one symbol at each time instant the vector ${\bf D}$ has this
structure which means that only one output of the RNN is used to
yield the symbol estimate. The initial symbol estimates for each
user (including the HD users and the LD users which use an MMSE
linear receiver \cite{verdu}) are then stacked in order to form
the $K \times 1$ vector with the initial decisions $\hat{\bf b}(i)
= [ \hat{b}_1(i) ~ \hat{b}_2(i) \ldots \hat{b}_{K}(i) ]$. The
estimate of the desired symbol for user $k$ at symbol $i$ then
becomes:
\begin{equation}
z_{k}(i) = {\bf D} \varphi({\bf W}_{k}^{H}(i)\boldsymbol{
\xi}_{k}(i))-{\bf f}_{k}^{H}(i)\hat{\bf b}(i)~,
~~~~~~k=1,2,\ldots,K_h
\end{equation}
where the $K \times 1$ vector ${\bf f}_{k}$ ($k=1, 2, \ldots,
K_h$) represents the feedback filters used to cancel the
interference contribution of the associated users. In particular,
the feedback filter ${\bf f}_{k}(i)$ of user $k$ has a number of
non-zero coefficients corresponding to the available number of
feedback connections for each type of cancellation structure. The
final detected QPSK symbol is:
\begin{equation}
\begin{split}
\hat{b}_{k}^{f}(i) & = {\rm sgn}\Big(\Re\Big[z_{k}(i)\Big]\Big) +
\j{\rm sgn}\Big(\Im\Big[z_{k}(i)\Big]\Big) \\  &
 = {\rm sgn}\Big(\Re\Big[ {\bf D} \varphi({\bf W}_{k}^{H}(i)\boldsymbol{
\xi}_{k}(i))-{\bf f}_{k}^{H}(i)\hat{\bf b}(i)\Big]\Big) + \j {\rm
sgn}\Big(\Im\Big[ {\bf D} \varphi({\bf W}_{k}^{H}(i)\boldsymbol{
\xi}_{k}(i))-{\bf f}_{k}^{H}(i)\hat{\bf b}(i)\Big]\Big)
\end{split}
\end{equation}

Regarding the structure of the decision feedback (DF) section, the
literature reports two basic types of feedback, namely, the
successive DF \cite{varanasi} and the parallel DF \cite{woodward}.
For successive DF (S-DF) \cite{varanasi2} and a system with an
equal number of feedback entries and users $K$, we have the
$K\times K$ matrix ${\bf F}(i)=[ {\bf f}_{1}(i)~\ldots~{\bf
f}_{K_h}(i)]$, which is formed with the filters for the $K$ users,
is strictly lower triangular, whereas for parallel DF (P-DF)
\cite{woodward} ${\bf F}(i)$ is full and constrained to have zeros
on the main diagonal in order to avoid cancelling the desired
symbols. The S-DF structure is optimal in the sense of that it
achieves the sum capacity of the synchronous CDMA channel with
AWGN \cite{varanasi2}. In addition, the S-DF scheme is less
affected by error propagation although it generally does not
provide uniform performance over the user population, which is a
desirable characteristic for uplink scenarios. In this context,
the P-DF system can offer uniform performance over the users but
is it suffers from error propagation in low signal-to-noise ratios
(SNRs) and high bit error rates (BERs). In order to design the DF
receivers and satisfy the constraints of the associated
structures, the designer must obtain the vector with initial
decisions $\bar{\bf b}(i)$ and then perform to the adopted
cancellation approach.

Unlike prior work, we focus on a scenario where there are $K_h$ HD
users, i.e., $K_h$ feedback filters ${\bf f}_k(i)$ with dimensions
$K \times 1$ as the filters are also supposed to cancel the
contribution of the LD users. Thus, in this case the matrix of
cancellation filters ${\bf F}(i)= [{\bf f}_1(i) ~{\bf f}_2(i)
~\ldots ~{\bf f}_{K_h}(i)]^T$ has dimensions $K \times K_h$. Due to
its property of offering uniform performance over the user
population and better performance as compared to S-DF schemes
\cite{woodward,delamare_mber,delamare_spadf}, we will adopt P-DF
schemes and introduce a modification for including the LD users in
the cancellation process. The feedback connections used and their
associated number of non-zero filter coefficients in ${\bf f}_{k}$
are equal to $K-1$ for the HD users. The resulting matrix ${\bf
F}(i)$ for the proposed neural receiver is full and constrained to
have zeros on the diagonal from the first row to the $K_h$th row in
order to avoid cancelling only the desired symbols of the HD users.
The rest of the matrix is full such that all LD are addressed by the
feedback section.

\section{Set-Membership Adaptive Algorithms}

Set-membership filtering (SMF) \cite{huang,gollamudi} represents a
class of recursive estimation algorithms that, on the basis of a
pre-determined error bound, seeks a set of parameters that yield
bounded filter output errors. These algorithms have been used in a
variety of applications such as adaptive equalization
\cite{gollamudi} and multi-access interference suppression
\cite{gollamudi}. The SMF algorithms are able to combat
conflicting requirements of fast convergence and low misadjustment
by introducing a modification on the objective function. In
addition, these algorithms exhibit reduced complexity due to
data-selective updates, which involve two steps: a) information
evaluation and b) update of parameter estimates. If the filter
update does not occur frequently and the information evaluation
does not involve much computational complexity, the overall
complexity can be significantly reduced. The adaptive SMF
algorithms usually achieve good convergence and tracking
performance due to an adaptive step size for each update, and
reduced complexity resulting from data selective updating.

In this section, a novel efficient parameter estimator for neural
structures based on the SMF framework \cite{huang,gollamudi} is
described. Unlike prior work on FIR filters, we propose
set-membership (SM) algorithms for RNNs structures. In the SM
framework, the receiver weight vectors ${\bf w}_{k,j}(i)$ and
${\bf f}_{k}(i)$ are designed to achieve a specified bound on the
magnitude of the estimation error $e_{k}(i)=b_{k}(i) - ({\bf
Dx}_{k}(i)- {\bf f}_{k}^{H}(i)\hat{\bf b}(i))$. As a result of
this constraint, the SM adaptive algorithm will only perform
filter updates for certain data. Let $\Theta_k (i)$ represent the
set containing all ${\bf w}_{k,j}(i)$ and ${\bf f}_{k}(i)$ that
yield an estimation error upper bounded in magnitude by $\gamma$.
Thus, we can write
\begin{equation}
\Theta_k(i) = \bigcap_{{({\bf u}(i),b_{k}(i))\in \mathbf{S}}} \{
{\bf w}_{k,j}, \in {\mathcal{C}}^{(Q+K_h)}, ~{\bf f}_{k} \in
{\mathcal{C}}^{K}:\mid e_{k}(i)\mid \leq \gamma \}
\end{equation}
where $\mathbf{S}$ is the set of all possible data pairs $({\bf
u}(i),b_{k}(i))$ and the set $\Theta_k(i)$ is referred to as the
feasibility set and any point in it is a valid estimate ${
z}_{k}(i)$. Since it is not practical to predict all data pairs,
adaptive methods work with the membership sets $\psi_{k,i}=
\bigcap_{m=1}^{i}  {\mathcal{H}}_{k,m}$ provided by the
observations, where ${\mathcal{H}}_{k,m}=\{{\bf w}_{k,j}, \in
{\mathcal{C}}^{(Q+K_h)}, ~{\bf f}_{k} \in {\mathcal{C}}^{K} :
|b_{k}(m) - z_{k}(m)| \leq \gamma\}$. In order to derive an SM
adaptive algorithm using point estimates for the proposed DF
receiver structure, we consider the cost function
\begin{equation}
{\mathcal{L}}({\bf W}_{k}(i),{\bf f}_{k}(i))=E[|b_{k}(i) - ({\bf
Dx}_{k}(i)- {\bf f}_{k}^{H}(i)\hat{\bf b}(i))|^2 ]
\end{equation}
subject to $|e_k(i)|\leq \gamma$, where $e_{k}(i) = b_{k}(i) -
({\bf Dx}_{k}(i)- {\bf f}_{k}^{H}(i)\hat{\bf b}(i))$. A gradient
algorithm can be developed by computing the gradient terms with
respect to ${\bf w}_{k,j}$ and ${\bf f}_{k}$, and using their
instantaneous values. Firstly, we consider the first partial
derivative of ${\mathcal{L}}({\bf W}_{k}(i),{\bf f}_{k}(i))$ with
respect to the parameter vector ${\bf w}_{k,j}(i)$ with dimension
$(Q+K_h)\times1$, which forms the matrix ${\bf W}_{k}$
\begin{equation}
\begin{split}
\frac{\partial {\mathcal{L}}({\bf W}_{k}(i),{\bf
f}_{k}(i))}{\partial{\bf w}_{k,j}^{*}(i)} & = \Bigg(\frac{\partial
e_{k}(i)}{\partial {\bf w}_{k,j}^{*}(i)}\Bigg) e_{k}^{*}(i)  = -
{\bf D} \Bigg(\frac{\partial {\bf x}_{k}(i)}{\partial {\bf
w}_{k,j}^{*}(i)}\Bigg) e_{k}^{*}(i) \\ & = - {\bf D}
\boldsymbol{\Lambda}_{k,j}(i) e_{k}^{*}(i)
\end{split}
\end{equation}
where the $K_h\times (Q+K_h)$ matrix
$\boldsymbol{\Lambda}_{k,j}(i)$ contains the partial derivatives
of the state vector ${\bf x}_{k}(i)$ with respect to ${\bf
w}_{k,j}(i)$. To obtain the expressions for updating the matrix
$\boldsymbol{\Lambda}_{k,j}(i)$, we consider the update equations
for the state vector ${\bf x}_{k}(i)$ given through (7) and (8).
Using the chain rule of calculus in (8), we obtain a recursion
that describes the dynamics of the learning process of the neural
section of the receiver:
\begin{equation}
\boldsymbol{\Lambda}_{k,j}(i+1) = \boldsymbol{\Phi}_{k}(i)\Big(
{\bf W}_{k}^{1:K}(i) \boldsymbol{\Lambda}_{k,j}(i) + {\bf
U}_{k,j}(i)\Big), ~j=1,\ldots,K
\end{equation}
where the $Q \times Q$ matrix ${\bf W}_{k}^{1:K}$ denotes the
submatrix of ${\bf W}_{k}$ formed by the first $K_h$ rows of ${\bf
W}_{k}$, the $Q \times Q$ matrix $\boldsymbol{\Phi}_{k}(i)$ for
user $k$ has a diagonal structure where the elements correspond to
the partial derivative of the activation function  $\varphi(.)$
with respect to the argument ${\bf
w}_{k,j}^{H}(i)\boldsymbol{\xi}_{k}(i)$ as given by
\begin{equation}
\boldsymbol{\Phi}_{k}(i) = diag\Big( \varphi'({\bf
w}_{k,1}^{H}(i)\boldsymbol{\xi}_{k}(i)), \ldots, \varphi'({\bf
w}_{k,K}^{H}(i)\boldsymbol{\xi}_{k}(i))\Big)
\end{equation}
where the $K_h\times (Q+K_h)$ matrix ${\bf U}_{k,j}(i)  = [ {\bf
0}^{T}~ \boldsymbol{\xi}_{k}^{T}(i)~ {\bf 0}^{T} ]^{T}$ has all
the rows with zero elements, except for the $j$-th row that is
equal to the vector $\boldsymbol{\xi}_{k}(i)$. By using a gradient
optimization that uses the rules ${\bf w}_{k,j}(i+1) = {\bf
w}_{k,j}(i) - \mu_{n} \frac{\partial {\mathcal{L}}({\bf
W}_{k}(i),{\bf f}_{k}(i))}{\partial{\bf w}_{k,j}^{*}(i)}$, ${\bf
f}_{k}(i+1) = {\bf f}_{k}(i) - \mu_{f} \frac{\partial
{\mathcal{L}}({\bf W}_{k}(i),{\bf f}_{k}(i))}{\partial{\bf
f}_{k}^{*}(i)}$ and substituting (14) and $\frac{\partial
{\mathcal{C}}({\bf W}_{k}(i),{\bf f}_{k}(i))}{\partial{\bf
f}_{k}^{*}(i)} = e_k^*(i)\hat{\bf b}(i)$, we obtain
\begin{equation}
{\bf w}_{k,j}(i+1) = {\bf w}_{k,j}(i) + \mu_{n} {\bf D}
\boldsymbol{\Lambda}_{k,j}(i)  e_{k}^{*}(i)
\end{equation}
\begin{equation}
{\bf f}_{k}(i+1) = {\bf f}_{k}(i) - \mu_{f} \hat{\bf b}(i)
\end{equation}
In order to ensure the constraint on the error bound ($|e_k(i)|
\leq \gamma$), we substitute (17) into (13) and then we make the
terms equal to $\gamma$. We proceed with the feedback filter ${\bf
f}_k(i)$ in the same way, which yields the algorithms
\begin{equation}
{\bf w}_{k,j}(i+1) = \left\{ \begin{array}{ll} { {\bf w}_{k,j}(i)
+ \frac{(1-\frac{\gamma}{|e_{k}(i)|})}
{||\boldsymbol{\Lambda}_{k,j}(i)||^{2}} {\bf D}
\boldsymbol{\Lambda}_{k,j}(i)
 e_{k}^{*}(i) } &
{\textrm{if $|e_{k}(i)|>\gamma$}} \\
{\bf w}_{k,j}(i) & \textrm{otherwise}\\
\end{array}\right.
\end{equation}
\begin{equation}
{\bf f}_{k}(i+1) = \left\{ \begin{array}{ll} {\bf f}_{k}(i) -
\frac{(1-\frac{\gamma}{|e_{k}(i)|})}{(\hat{\bf b}^{H}(i){\bf
b}(i))}  e_{k}^{*}(i) \hat{\bf b}(i) &
{\textrm{if $|e_{k}(i)|>\gamma$}} \\
{\bf f}_{k}(i)  & \textrm{otherwise}\\
\end{array}\right.
\end{equation}
where $\mu_{n}=(1-\frac{\gamma}{|e_{k}(i)|})
/||\boldsymbol{\Lambda}_{k,j}(i)||^{2}$ and
$\mu_{f}=(1-\frac{\gamma}{|e_{k}(i)|})/(\hat{\bf b}^{H}(i){\bf
b}(i))$ are the new normalized step sizes of the algorithms. The
novel adaptive technique is called the SM normalized real time
recurrent learning (SM-NRTRL) algorithm. It introduces a powerful
variable step size and discerning update rule.

\section{Simulation Experiments}

In this section we assess the performance of the proposed
space-time neural receivers and SM algorithms in terms of the bit
error rate (BER). We compare the proposed neural MUD (NMUD) and DF
neural MUD (DF-NMUD) receivers with the RAKE \cite{paulraj}, the
linear MUD (L-MUD) and the DF-MUD with $J=1$ and their space-time
versions with $J=2,3$ antenna elements. We also evaluate the
convergence performance of the adaptive estimation algorithms,
i.e., the NLMS, the NRTRL \cite{haykin-neural}, the SM-NLMS
\cite{huang} and the proposed SM-NRTRL.  For the NMUD we make
${\bf f}_{k}={\bf 0}$ in the structure and algorithms of Sections
III and IV.

The DS-CDMA system employs, QPSK modulation, Gold sequences of
length $N=7$ or random sequences of length $N=8$ for the HD users
and Gold sequences of length $N_l=31$ for the LD users and the
number of HD and LD users is equal ($K_l=K_h$) in all experiments.
We adopted an identical ratio of LD to HD users for simplicity. In
a practical system, it is expected that the number of LD users
significantly exceed the HD ones since the LD users correspond to
voice users (major part of users and market), whereas the HD users
correspond to multimedia and data users. Another issue is that
since a number of DS-CDMA systems in use include higher-order
modulation schemes, besides QPSK, it should be noted that the
proposed system can be modified to handle higher-order modulation
schemes such as QAM-16 and QAM-64. In this case, it would be
necessary to modify the detection rule in the receiver.
Specifically, the slicer would have to take into account the
different levels of the signal constellation. The LD users also
operate asynchronously with single-antenna MMSE linear receivers
\cite{paulraj} and a power level $3$ dB below that of HD users.
The channels experienced by different users are i.i.d. whose
coefficients for each user $k$ ($k=1,\ldots,K$) are
$h_{k,l}(i)=p_{l}\alpha_{k,l}(i)$, where $p_l$ are the gains (or
average powers) of the channel paths (power delay profile) and
$\alpha_{k,l}(i)$ ($l=0,1,\ldots,L-1$) are complex Gaussian random
sequences obtained by passing complex white Gaussian noise through
a filter with approximate transfer function
$c/\sqrt{1-(f/f_{d})^{2}}$ where $c$ is a normalization constant,
$f_{d}=v/\lambda$ is the maximum Doppler shift, $\lambda$ is the
wavelength of the carrier frequency, and $v$ is the speed of the
mobile \cite{rappa}. The channel powers are normalized so that
$\sum_{l=1}^{L_{p}}p_{l}^{2}=1$. This procedure corresponds to the
generation of independent sequences of correlated unit power
Rayleigh random variables ($E[|\alpha ^2_{k,l}(i)| ]=1$). We show
the results in terms of the normalized Doppler frequency $f_{d}T$
(cycles/symbol) and use three-path channels with relative average
powers $p_l$ given by $0$, $-3$ and $-6$ dB, where in each run the
spacing between paths is obtained from a discrete uniform random
variable (d.u.r.v.) between $1$ and $3$ chips.

The asynchronism $\tau_k$ in chips is taken from a d.u.r.v between
$1$ and the respective processing gain of the class of user (HD or
LD). The parameters of the algorithms are optimized and the system
has perfect power control between users of each class (HD or LD).
The activation function $\varphi(.)$ for the neural receivers is
the hyperbolic tangent, the number of states $Q$ of the RNNs is
set to $1$, the DOAs $\phi_{k,m}$ are uniformly distributed in
$(-\pi/3,\pi/3)$ for all simulations and the NLMS is used to
estimate the effective spatial signature of the receiver
structure. Experiments are averaged over $200$ independent runs
and the parameter $\gamma$ was chosen as $\sqrt{3.5 \sigma^2}$ for
the neural MUDs and $\sqrt{5 \sigma^2}$ for the other detectors.
The proposed and existing algorithms are all initialized with
parameter vectors with zero elements in order to provide a fair
comparison. The BER performance shown in the results refers to the
average BER amongst the $K_h$ HD users.

In Figs. 3 and 4 we show the BER convergence performance of the
analyzed algorithms and linear and DF receivers, respectively. We
also include the Kalman algorithm of \cite{choi} with $J=3$ and
the proposed neural structure in the comparison as an upper bound.
The curves show that data-selective algorithms outperform the
conventional adaptive techniques both in convergence speed and
steady-state performance, while they only perform filter updates
for a fraction of the overall symbols. The DF schemes, shown in
Fig. 4, and in particular the DF-NMUD outperform the schemes
without DF , shown in Fig. 3, and as the number of antenna
elements $J$ is increased so is the performance. The proposed
SM-NRTRL algorithm for the neural structure outperforms the NRTRL
and has a performance very close to the Kalman technique of
\cite{choi}, which has much higher complexity ($O(K_h^3)$).

In terms of complexity, the algorithms require $O(QK_h)$ for
training the neural structures and $O(K)$ for training the linear
filters but the SM-NRTRL neural algorithms exhibit better
performance and lower UR. Indeed, the update rate (UR) of the SM
algorithms is shown in Table I at $E_{b}/N_{0}=10 dB$ and
$f_{d}T=0.0001$, where it is shown the advantage in terms of UR of
the neural approach combined with SM framework. The UR depends on
$f_{d}T$ and $\gamma$, and we verified that, for an extensive set
of scenarios, the SM-NRTRL has the ability to consistently operate
at lower UR than existing algorithms. The complexity is mainly
dictated by the number of states $Q$ of the neural network and the
number of HD users $K_h$. The number of antenna elements $J$
increases the complexity of the space-time receivers in an
identical way for both neural and non-neural schemes because the
in the proposed structure the signal processing either via neural
structures or linear ones is performed after the bank of ST-RAKE
receivers. In this case, we used NLMS algorithms to estimate the
effective spatial signatures of the users, which despread the
signals prior to the processing of the receiver.

The BER performance of SM-adaptive algorithms and DF receivers is
shown in Fig. 5. The receivers are adjusted with $200$ symbols
during the training period, then switch to decision-directed mode
and process $2000$ data symbols. The results show that the
proposed DF-NMUD achieves the best BER performance, followed by
the DF-MUD with linear filters (DF-MUD), and the RAKE receivers.

In the last experiment, shown in Fig. 6, we consider randomly
generated sequences with $N=8$ and receivers with single antennas
and space-time receivers with $J=3$ antenna elements in order to
evaluate a situation that may arise in a practical standard, i.e.,
the 3G UMTS system. In this case, the channels are modeled by the
UMTS Vehicular A channel model \cite{umts} with maximum Doppler
frequency $f_D = 100$ Hz with a carrier frequency $f_c=2$ GHz. The
receivers are adjusted with $200$ symbols during the training
period, then switch to decision-directed mode and process $2000$
data symbols. The curves show that the proposed DF-NMUD achieves
the best BER performance, followed by the remaining techniques,
corroborating the results obtained for the previous experiments.

\section{Conclusions and Future Work}

A space-time adaptive DF neural receiver for high rate users in
DS-CDMA systems was proposed for joint equalization and multiuser
detection. In order to estimate the parameters of the proposed
receiver, we developed a low complexity data-selective algorithm
was developed based on the concept of Set-Membership filtering.
Numerical results have shown that the proposed neural structure
and algorithms considerably outperform existing methods. For
future work, the proposed scheme and algorithms can also be
considered and applied to other communications systems such as
MIMO systems. Detectors with attractive trade-off between
performance and complexity such as the proposed one are of central
importance in MIMO systems. This would require a description of a
discrete-time model of MIMO systems and a formulation of the
problem involving the application of the proposed DF neural
structure for extracting the desired signals embedded in
interference. Future work may also involve the investigation of
novel adaptive estimation algorithms using the set-membership
framework. Algorithms with faster convergence rate than the one
proposed here such as least squares and those with data reuse may
be of interest for the estimation of parameters in a larger neural
structure, as well as, the formulation of set-membership
techniques with automatic adjustment of the bound.

\begin{figure}[!h]
\begin{center}
\def\epsfsize#1#2{1\columnwidth}
\epsfbox{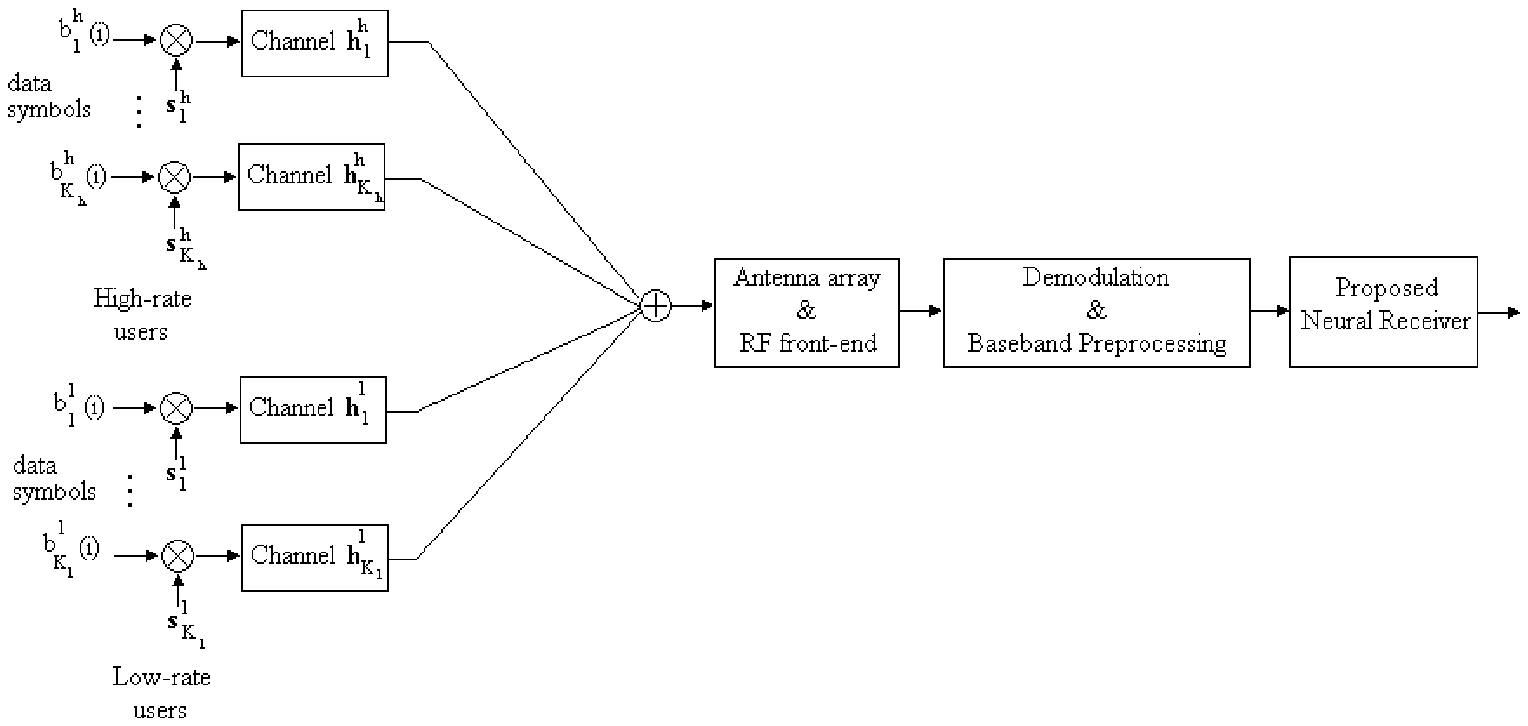} \caption{}
\end{center}
\end{figure}

\newpage
\begin{figure}[!htb]
\begin{center}
\def\epsfsize#1#2{0.9\columnwidth}
\epsfbox{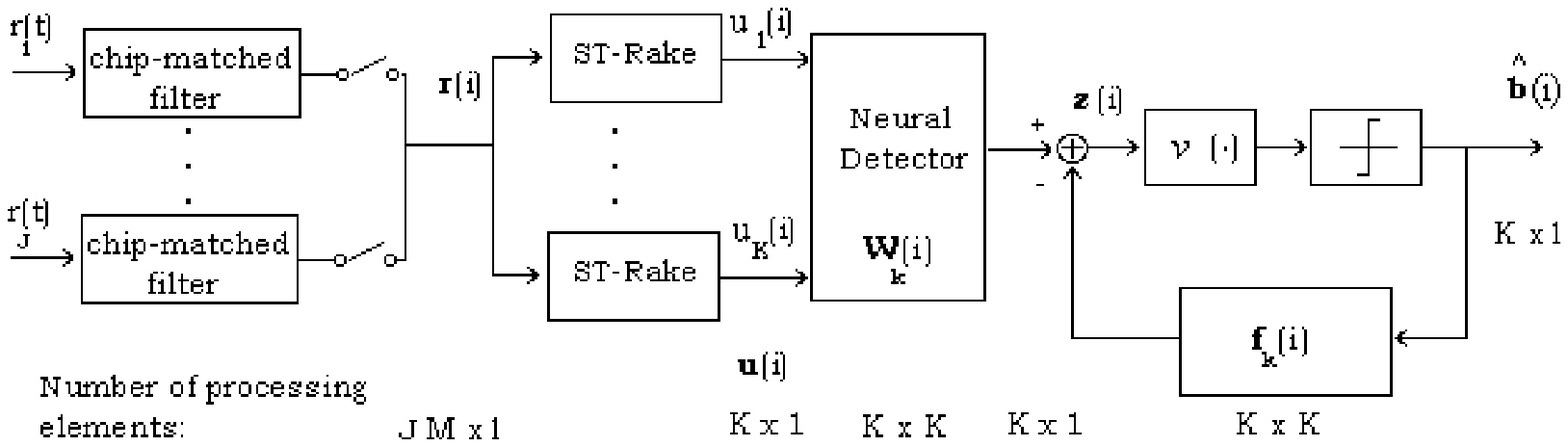} \caption{}
\end{center}
\end{figure}

\begin{figure}[!htb]
\begin{center}
\def\epsfsize#1#2{0.9\columnwidth}
\epsfbox{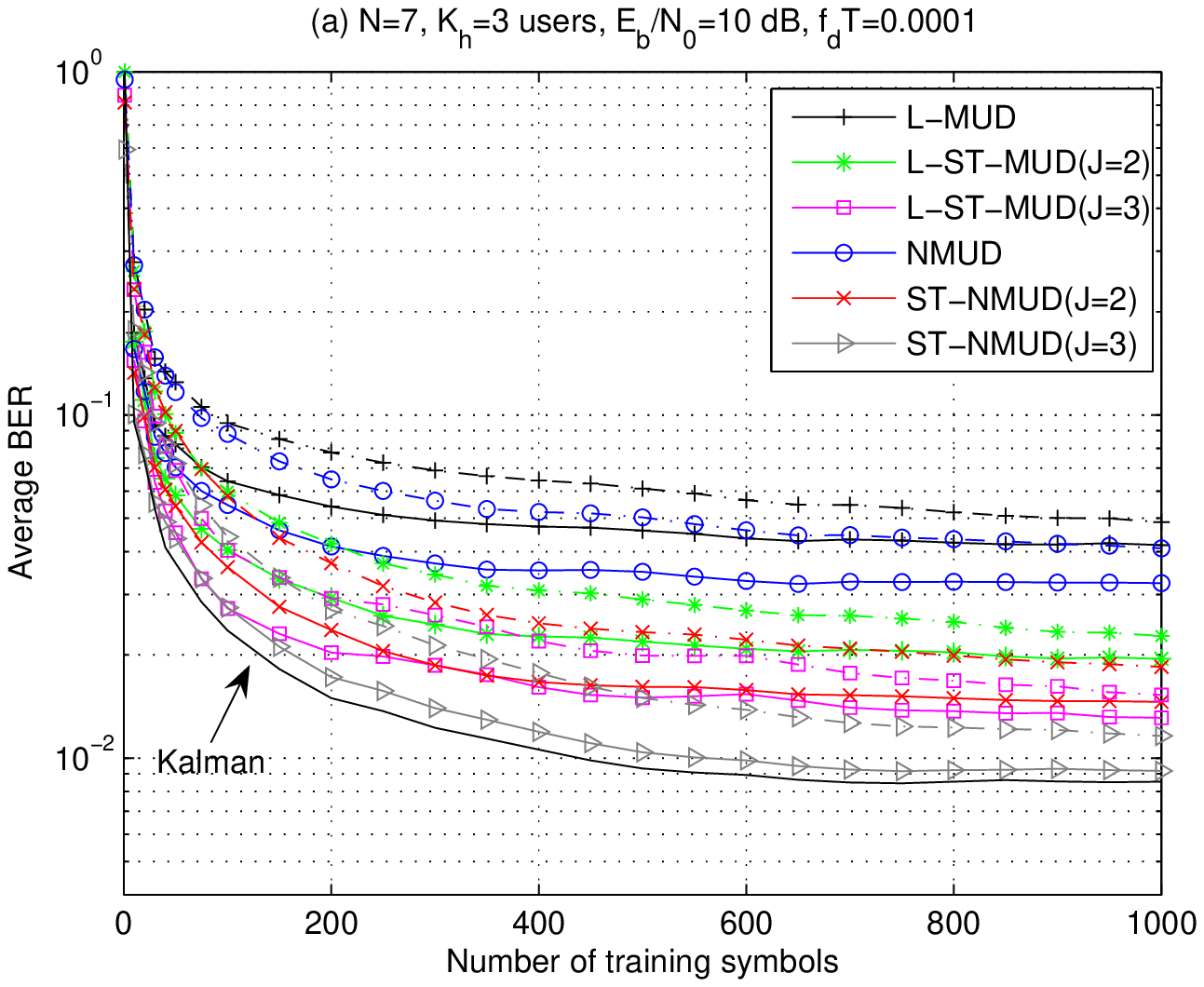} \caption{}
\end{center}
\end{figure}

\begin{figure}[!htb]
\begin{center}
\def\epsfsize#1#2{0.9\columnwidth}
\epsfbox{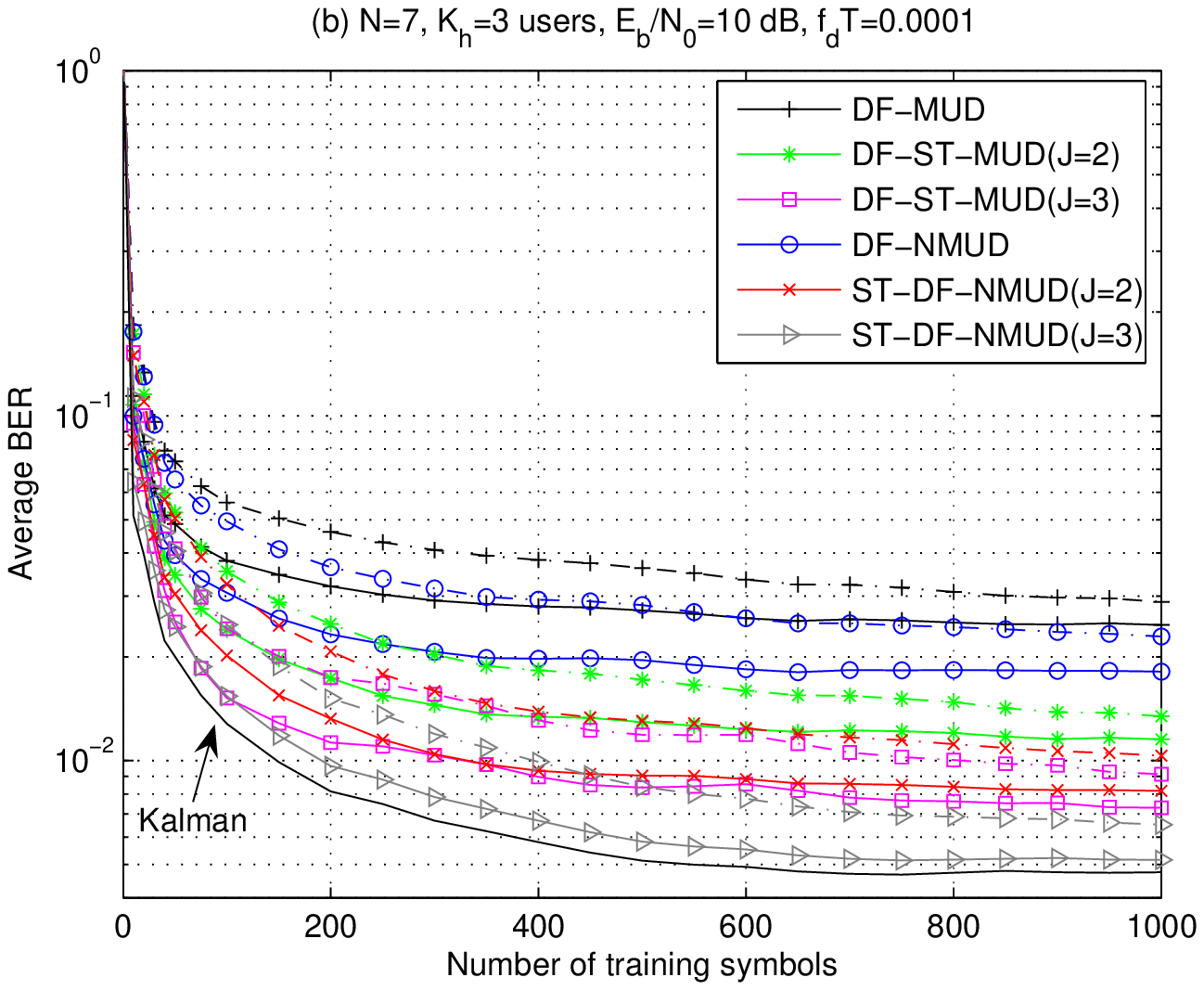} \caption{}
\end{center}
\end{figure}


{\linespread{1.8}

\begin{table}[h]
\centering
 \caption{}
\begin{tabular}{|c|c|c|c|c|}
  \hline
  { Detectors } & L-MUD & NMUD & DF-MUD & DF-NMUD \\
  \hline
  J=1 & {25.4$\%$}  & {18.2$\%$} & {24.6$\%$} & {15.7$\%$} \\
  \hline
  J=2 & {24.5$\%$}  & {17.1$\%$} & {23.1$\%$} & {14.3$\%$} \\
  \hline
  J=3 & {23.9$\%$}  & {16.8$\%$} & {21.5$\%$} & {13.8$\%$} \\
  \hline
\end{tabular}
\end{table}
}

\begin{figure}[!htb]
\begin{center}
\def\epsfsize#1#2{0.9\columnwidth}
\epsfbox{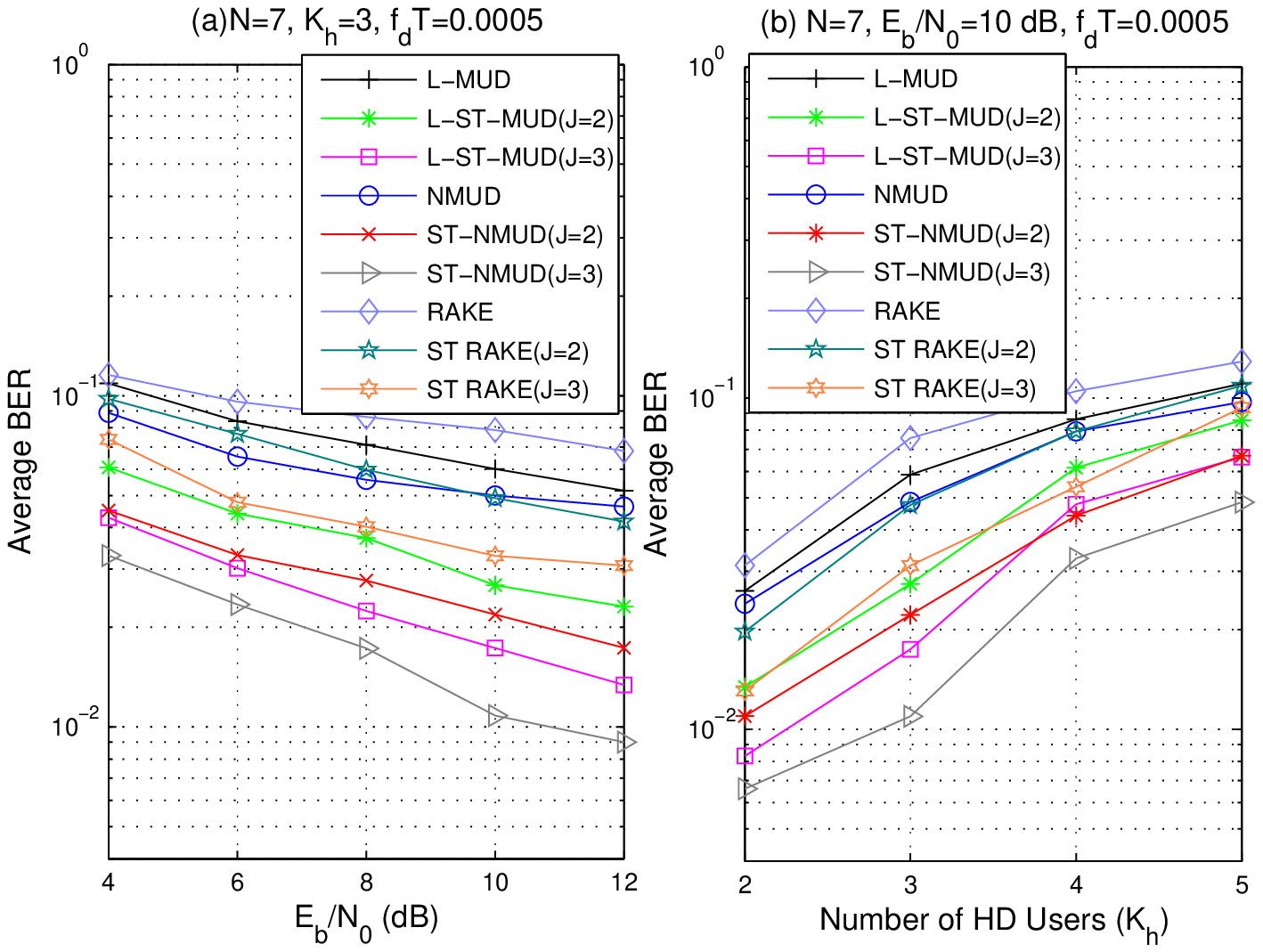} \caption{}
\end{center}
\end{figure}

\begin{figure}[!htb]
\begin{center}
\def\epsfsize#1#2{0.9\columnwidth}
\epsfbox{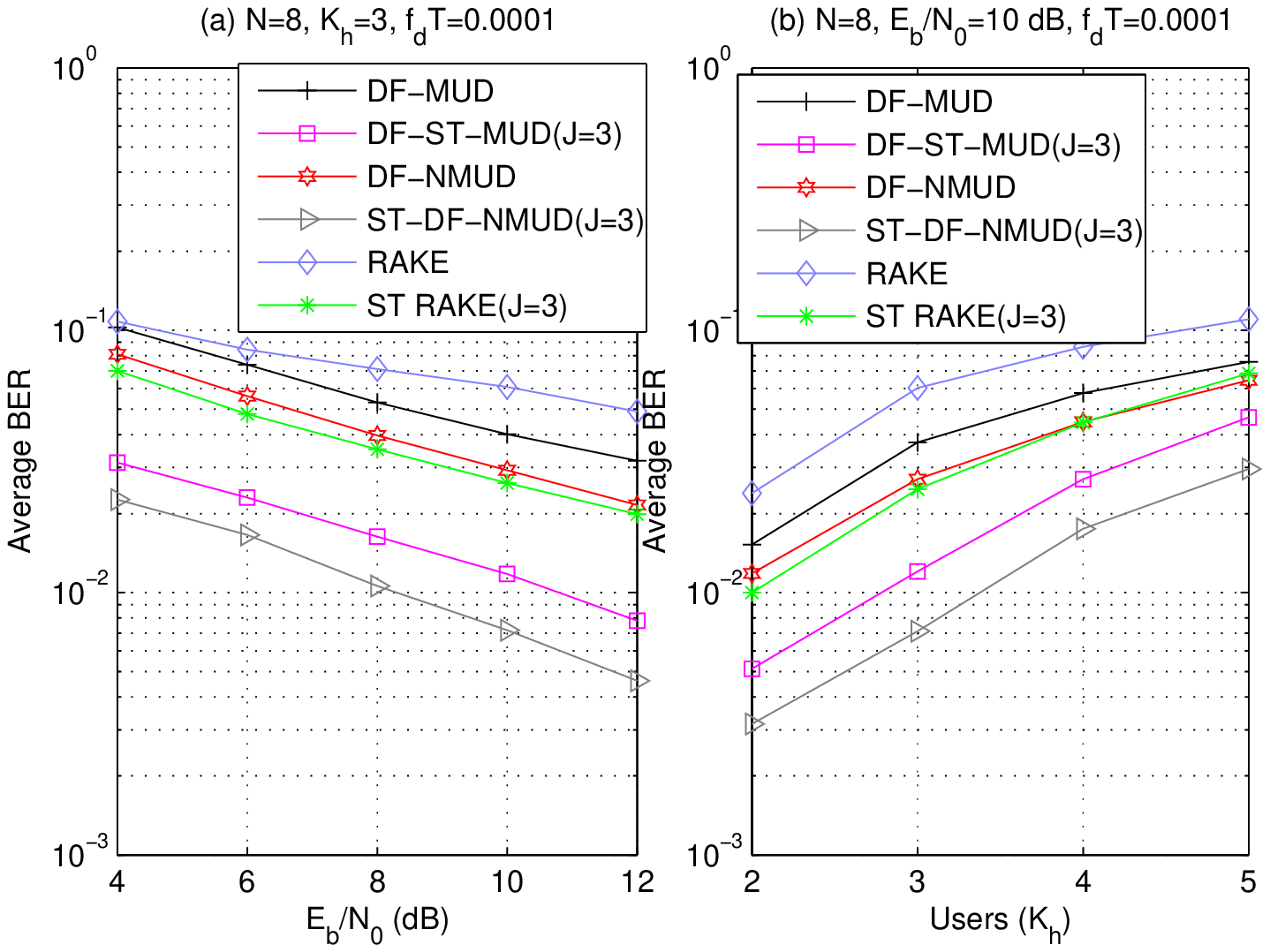} \caption{}
\end{center}
\end{figure}

{\bf Captions:}

Fig. 1: {Block diagram of the space-time DS-CDMA system model.}

Fig. 2: {Block diagram of the proposed space-time decision
feedback neural receiver.}

Fig. 3: {BER convergence performance of data-seletive and
conventional (dash-dotted lines) algorithms for linear receivers.}

Fig. 4: {BER convergence performance of data-seletive and
conventional (dash-dotted lines) algorithms for DF receivers.}

Table 1: {Update Rate of SM algorithms at $E_{b}/N_{0}=10 dB$ and
$f_{d}T=0.0001$.}

Fig. 5: {BER performance of DF receivers versus (a) $E_{b}/N_{0}$
and (b) number of HD users ($K_h$).}

Fig. 6: {BER performance of DF receivers versus (a) $E_{b}/N_{0}$
and (b) number of HD users ($K_h$) for a UMTS channel model.}

\end{document}